\documentclass[aps,prd,twocolumn,showpacs]{revtex4}

\usepackage{graphicx}
\usepackage{dcolumn}
\usepackage{bm}
\usepackage{simplewick}

\begin{document}

\title{Probability distribution of the vacuum energy density}

\author{Goran Duplan\v ci\' c$^{a}$}
\email{gorand@thphys.irb.hr}
\author{Dra\v zen Glavan$^{b}$}
\email{dglavan@dominis.phy.hr}
\author{Hrvoje \v Stefan\v ci\' c$^{a}$}
\email{shrvoje@thphys.irb.hr}
\affiliation{$^{a}$Theoretical Physics Division,
Rudjer Boskovic Institute,
P.O.Box 180,
HR-10002 Zagreb,
Croatia
}
\affiliation{$^{b}$Department of Physics, 
Faculty of Science, University of Zagreb, 
P.O.Box 331
HR-10002 Zagreb,
Croatia}

\date{\today}

\begin{abstract}
As the vacuum state of a quantum field is not an eigenstate of 
the Hamiltonian density, the vacuum energy density can be represented as 
a random variable. We present an analytical calculation of the probability 
distribution of the vacuum energy density for real and complex massless 
scalar fields in Minkowski space. 
The obtained probability distributions are broad and the 
vacuum expectation value of the Hamiltonian density is not fully 
representative of the vacuum energy density. 
\end{abstract}


\maketitle

Quantum field theory (QFT) is a theoretical framework that has enabled, 
via its phenomenologically successful models such as the Standard Model, 
a detailed quantitative understanding of a vast number of physical phenomena. 
In numerous areas of its application QFT predictions are confirmed at ever 
greater accuracy with every new experiment. 
Yet, these QFT models with such a remarkable performance in precise 
determination of scattering cross-sections and decay rates also give 
a definite prediction for the vacuum expectation value (VEV) of the 
Hamiltonian density. As long as we disregard gravitational interaction, 
the size of this VEV is irrelevant
\cite{BjorkenDrell}. When the 
gravitational interaction, described by general relativity, is included
the VEV of the Hamiltonian density can no longer be ignored. Moreover, any 
estimate of this quantity within QFT, also referred to as 
{\em zero point energy density}, shows that this contribution to the 
source term of the Einstein equation is crucial. The zero point energy 
density represents a major contribution to the total value of the cosmological 
constant (CC) \cite{weinberg}. The fact that the size of this contribution, 
as well as the size of other identified contributions to the cosmological 
constant, are many orders of magnitude larger than the observed value of 
the CC is at the core of the notorious {\em old cosmological constant problem}. 
A deeper understanding of all contributions to the cosmological constant 
is clearly needed to resolve the old CC problem, perceived as one of the 
most serious problems in theoretical physics.
Another important physical phenomenon where the understanding of the energy 
density in the vacuum is paramount is {\em inflation} \cite{inf}.

In this paper our primary goal is understanding of the energy density that 
quantum fields develop in the vacuum state. Hereafter we call this energy 
density {\em the vacuum energy density} and investigate how well it is 
represented by a VEV of the Hamiltonian density.   
The central result of this paper is the calculation of the probability 
density function (p.d.f.) of the vacuum energy density for both real and 
complex free massless scalar fields in Minkowski space. We present a fully analytically 
tractable treatment of the problem and finally provide the analytical 
expressions for the probability density functions of the vacuum energy 
density. 
As our motivation comes from cosmology we do not assume normal ordering during
calculation. However, since normal ordering is a widely accepted regularization procedure in
QFT we present ``normally ordered'' results in the discussion.
Throughout the paper we adopt the notation and conventions
of \cite{BjorkenDrell}. 

We start with a direct observation that, for both real and complex scalar 
fields, the Hamiltonian $H$ and the Hamiltonian density ${\cal H}$ operators 
do not commute. The vacuum state $|0 \rangle$ is not an eigenstate of 
${\cal H}$. The measurement of the Hamiltonian density in vacuum can 
result in different values of vacuum energy density with different 
probabilities. 
A quick calculation of the variance $\sigma_{\cal H}^2=\langle {\cal H}^2 \rangle - \langle {\cal H} 
\rangle^2$ shows that $\sigma_{\cal H}^2= 2/3 \langle {\cal H} \rangle^2$ for real and 
$\sigma_{\cal H}^2= 1/3 \langle {\cal H} \rangle^2$ for complex massless scalar field.
The result that $\sigma_{\cal H}$ is of 
the same order as $\langle {\cal H} \rangle$ 
reveals the existence of a broad p.d.f. 
for the vacuum energy density. 
At this point, one can question the relevance of ${\cal H}$ as an observable.
Namely, from the practical point of view, it is more natural (i.e. physical) to observe a smoothed
energy density in a finite space-time region. In that case, the relevant region depends on the size
of the system under investigation. It can be as big as the horizon in the case of cosmology or as
small as experimental setup in the case of Casimir effect. The upper limit corresponding to the infinite region gives
an uninteresting result (p.d.f. is a delta function because ${\cal H} \propto H$) and the lower limit corresponds to
taking ${\cal H}$ as an observable.  Smoothing of energy density in a finite space-time region considerably complicates
the calculation. 
Here we assume that ${\cal H}$ is a legitimate observable and our result is directly useful as a limiting value for making
measurements in smaller and smaller volumes. It is important to note that treating ${\cal H}$ as an observable can be
crucial in understanding cosmological inflation where our 
patch of the Universe originates from a virtually pointlike element of space.

Since the spectrum of the Hamiltonian density operator is positively defined, the p.d.f. 
of the vacuum energy density is to be determined by using a Laplace transform. 
If $p(\epsilon)$ is the p.d.f. of interest, its Laplace transform is given by
\begin{eqnarray}
L\{p(\epsilon)\} &=&  \int_0^\infty e^{-s\,\epsilon}p(\epsilon)\,d\epsilon = 
\sum_{n=0}^{\infty}\frac{(-s)^n}{n!} \int_0^\infty \epsilon ^n \,p(\epsilon)\,
d\epsilon  \nonumber \\
 &=&  \sum_{n=0}^{\infty}\frac{(-s)^n}{n!} \langle {\cal H}^n \rangle , \label{transf}
\end{eqnarray}
where $\langle {\cal H}^n \rangle $ is the VEV of $n$-th power of the Hamiltonian density 
operator.
Dimensions of $s$ and $\epsilon$ are $[E]^{-4}$ and $[E]^4$, respectively.
Since the original function is uniquely defined by its image, once the transform 
$L\{p(\epsilon)\}$ is known, the p.d.f. $p(\epsilon)$ can be determined by using 
various properties of the Laplace transform
\cite{Abram}.

To obtain the transform $L\{p(\epsilon)\}$ one has to calculate 
$\langle {\cal H}^n \rangle $, $n=1,\ldots,\infty$.
Hamiltonian densities of real (${\cal H}_{\rm R}$) and complex 
(${\cal H}_{\rm C} $) massless scalar fields are given by
\begin{equation}
{\cal H}_{\rm R}(\mathbf{x},t)  =  -\int d^3 k\,d^3 q \,\frac{\omega_\mathbf{k} 
\omega_\mathbf{q}+\mathbf{k}\cdot \mathbf{q}}{4 (2 \pi)^3 \sqrt{\omega_\mathbf{k}
 \omega_\mathbf{q}}} ~\, r_\mathbf{k}(\mathbf{x},t)\, r_\mathbf{q}(\mathbf{x},t), 
 \label{Hr} 
\end{equation}
\begin{equation}
{\cal H}_{\rm C}(\mathbf{x},t)  =  ~~\int d^3 k\,d^3 q \,\frac{\omega_\mathbf{k} 
\omega_\mathbf{q}+\mathbf{k}\cdot \mathbf{q}}{2 (2 \pi)^3 \sqrt{\omega_\mathbf{k}
 \omega_\mathbf{q}}} ~\, c_\mathbf{k}^\dagger(\mathbf{x},t)\, c_\mathbf{q}(\mathbf{x},t),
 \label{Hc}
\end{equation}
where $\omega_\mathbf{p}=|\mathbf{p}|$,
\begin{eqnarray}
r_\mathbf{p}(\mathbf{x},t)& = & a_\mathbf{p} 
e^{-i(\omega_\mathbf{p}\,t-\mathbf{p}\cdot\mathbf{x})}- a^\dagger_\mathbf{p}
e^{i(\omega_\mathbf{p}\,t-\mathbf{p}\cdot\mathbf{x})}, \label{rp} \\
c_\mathbf{p}(\mathbf{x},t)& = & a_\mathbf{p} 
e^{-i(\omega_\mathbf{p}\,t-\mathbf{p}\cdot\mathbf{x})}- b^\dagger_\mathbf{p}
e^{i(\omega_\mathbf{p}\,t-\mathbf{p}\cdot\mathbf{x})} \label{rc} ,
\end{eqnarray}
and $a^\dagger_\mathbf{p}$, $b^\dagger_\mathbf{p}$ ($a_\mathbf{p}$, $b_\mathbf{p}$) are 
the usual creation (annihilation) operators of scalar fields.
For further considerations it is convenient to know contractions and commutators of the 
above mentioned operators. The relevant contractions and commutators are
\begin{eqnarray}
& & \contraction{}{r}{_{\mathbf{k}}~}{r}
r_{\mathbf{k}}~ r_{\mathbf{q}} = -\delta ^3(\mathbf{k}-\mathbf{q})\, 
e^{-i(\omega_\mathbf{k}-\omega_\mathbf{q})\, t+i(\mathbf{k}-\mathbf{q})\cdot \mathbf{x}},\label{cont0}\\
& & \contraction{}{c}{_{\mathbf{k}}~}{c} c_{\mathbf{k}}~ c^\dagger_{\mathbf{q}}=
\contraction{}{c}{^\dagger_{\mathbf{k}}~}{c}
c^\dagger_{\mathbf{k}}~ c_{\mathbf{q}}=\delta ^3(\mathbf{k}-\mathbf{q})\,  
e^{-i(\omega_\mathbf{k}-\omega_\mathbf{q})\, t+i(\mathbf{k}-\mathbf{q})\cdot \mathbf{x}},\label{cont1}\\
& & \contraction{}{c}{_{\mathbf{k}}~}{c} c_{\mathbf{k}}~ c_{\mathbf{q}}=
\contraction{}{c}{^\dagger_{\mathbf{k}}~}{c} c^\dagger_{\mathbf{k}}~ c^\dagger_{\mathbf{q}}=0, \label{cont2} \\
& & {}[r_{\mathbf{k}},r_{\mathbf{q}} ]= [c_{\mathbf{k}},c_{\mathbf{q}}] = [c^{\dagger}_{\mathbf{k}},c^{\dagger}_{\mathbf{q}}]
=[c_{\mathbf{k}},c^{\dagger}_{\mathbf{q}}]=0\, , \label{comm}
\end{eqnarray}
where, from now on, arguments $(\mathbf{x},t)$ of the operators $r$ and $c$ are omitted 
for the sake of notation simplicity. 

From Eqs. (\ref{Hr}) and (\ref{Hc}), it follows that vacuum expectation value of ${\cal H}^n $ can be written as
\begin{eqnarray}
\langle {\cal H}^n_{\rm R} \rangle  & = &  \int \frac{d^3 k_1\ldots d^3 k_{2n}}{4^n (2 \pi)^{3n}} 
\prod_{i=1}^{n}\frac{\omega_{\mathbf{k}_i} 
\omega_{\mathbf{k}_{n+i}}+{\mathbf{k}_i}\cdot {\mathbf{k}_{n+i}}}{\sqrt{\omega_{\mathbf{k}_i}
 \omega_{\mathbf{k}_{n+i}}}} \nonumber \\
& & {}\times  (-)^n  \langle 0|\prod_{j=1}^{n} r_{\mathbf{k}_j}\, r_{\mathbf{k}_{n+j}} |0\rangle ,~
\label{vevHrn} \\
\langle {\cal H}^n_{\rm C} \rangle  & = &  \int \frac{d^3 k_1\ldots d^3 k_{2n}}{2^n (2 \pi)^{3n}} 
\prod_{i=1}^{n}\frac{\omega_{\mathbf{k}_i} 
\omega_{\mathbf{k}_{n+i}}+{\mathbf{k}_i}\cdot {\mathbf{k}_{n+i}}}{\sqrt{\omega_{\mathbf{k}_i}
 \omega_{\mathbf{k}_{n+i}}}} \nonumber \\
& & {}\times  \langle 0|\prod_{j=1}^{n} c^\dagger_{\mathbf{k}_j}\, c_{\mathbf{k}_{n+j}} |0\rangle ,~
\label{vevHcn}
\end{eqnarray}
where the notation for momenta is renamed for later convenience.

The expectation values in integrands of Eqs. (\ref{vevHrn}) and (\ref{vevHcn}) can be calculated by 
using Wick's theorem \cite{Wick} and contractions from 
Eqs. (\ref{cont0}) -- (\ref{cont2}). It follows that
\begin{eqnarray}
\lefteqn{\langle 0|\prod_{i=1}^{n} r_{\mathbf{k}_i}\, r_{\mathbf{k}_{n+i}} |0\rangle  
 = \frac{(-)^n}{2^n n!}  } \label{r} \\
& & {}
{{{}\atop {\displaystyle\sum\nolimits}}\atop {{\{i_1,\ldots,i_{2n}\}=P\{1,\ldots,2n\}}\atop {}}} 
 \delta(\mathbf{k}_{i_1}-\mathbf{k}_{i_2})
\ldots \delta(\mathbf{k}_{i_{2n-1}}-\mathbf{k}_{i_{2n}}) ,\nonumber 
\end{eqnarray}
\begin{eqnarray}
\lefteqn{\langle 0|\prod_{i=1}^{n} c^\dagger_{\mathbf{k}_i}\, c_{\mathbf{k}_{n+i}} |0\rangle = }\label{c} \\ 
& & {}
{{{} \atop {\displaystyle\sum\nolimits}} 
\atop {{\{j_1,\ldots,j_{n}\}=P\{n+1,\ldots,2n\}} \atop{} }}\,  
\delta(\mathbf{k}_{1}-\mathbf{k}_{j_1}) 
\ldots \delta(\mathbf{k}_{n}-\mathbf{k}_{j_{n}}) ,\nonumber
\end{eqnarray}
where $P$ are permutations. 
The total numbers of terms in Eqs. (\ref{r}) and (\ref{c}) are $(2n)!/(2^n n!)$ 
and $n!$, respectively.

Because of the symmetry on the exchange of dummy variables $\mathbf{k}_i$, it is possible 
to further simplify  Eqs. (\ref{vevHrn}) and (\ref{vevHcn}) using binomial coefficients
\begin{eqnarray}
\lefteqn{\langle {\cal H}^n_{\rm R} \rangle   =   \sum_{l=0}^n \left( n \atop l \right)\int 
\frac{d^3 k_1\ldots d^3 k_{2n}}{4^n (2 \pi)^{3n}}
\left(\prod_{i=1}^l \sqrt{\omega_{\mathbf{k}_i}
 \omega_{\mathbf{k}_{n+i}}}\right) } \nonumber \\
& &{}\times \left(\prod_{s=l+1}^{n} \frac{{\mathbf{k}_s}\cdot {\mathbf{k}_{n+s}}}{ \sqrt{\omega_{\mathbf{k}_s}
 \omega_{\mathbf{k}_{n+s}}}} \right)
(-)^n \langle 0|\prod_{j=1}^{n} r_{\mathbf{k}_j}\, r_{\mathbf{k}_{n+j}} |0\rangle ,~~~~{}
\label{vevHrn1a}  
\end{eqnarray}
\begin{eqnarray}
\lefteqn{\langle {\cal H}^n_{\rm C} \rangle   =  \sum_{l=0}^n \left( n \atop l \right)\int 
\frac{d^3 k_1\ldots d^3 k_{2n}}{2^n (2 \pi)^{3n}}
\left(\prod_{i=1}^l\sqrt{\omega_{\mathbf{k}_i}
 \omega_{\mathbf{k}_{n+i}}} \right) } \nonumber \\
& &{}\times \left(\prod_{s=l+1}^{n}  \frac{{\mathbf{k}_s}\cdot {\mathbf{k}_{n+s}}}{ \sqrt{\omega_{\mathbf{k}_s}
 \omega_{\mathbf{k}_{n+s}}}}\right)
 \langle 0|\prod_{j=1}^{n} c^\dagger_{\mathbf{k}_j}\, c_{\mathbf{k}_{n+j}} |0\rangle ,~
\label{vevHcn1a}
\end{eqnarray}
where we assume that in the case of an ill-defined product ($\prod _{i=1}^0$ and 
$\prod _{i=n+1}^n$) the whole contribution in corresponding brackets is equal to 1.

Since the factors ${\mathbf{k}_i}\cdot {\mathbf{k}_j}$ are odd functions of 
momenta and taking into account Eqs. (\ref{r}) and (\ref{c}),
the nonvanishing contributions in Eqs. (\ref{vevHrn1a}) and (\ref{vevHcn1a}) can be 
factorized. In factorized form Eqs.  (\ref{vevHrn1a}) and (\ref{vevHcn1a}) are given by
\begin{eqnarray}
\lefteqn{\langle {\cal H}^n_{\rm R} \rangle  =  }\label{vevHrn1ax} \\
& &{} \frac{1}{2^n} \sum_{l=0}^n \left( n \atop l \right)
\Bigg [\int \frac{d^3 k_1\ldots d^3 k_{l}\, d^3 k_{n+1}\ldots d^3 k_{n+l}}{2^l (2 \pi)^{3l}} \nonumber \\  
& &{}\times
\left(\prod_{i=1}^l \sqrt{\omega_{\mathbf{k}_i}
 \omega_{\mathbf{k}_{n+i}}}\right)
 (-)^l\langle 0|\prod_{j=1}^{l} r_{\mathbf{k}_j}\, r_{\mathbf{k}_{n+j}} |0\rangle \Bigg ]
 \nonumber \\
& &{}\times  \Bigg [\int \frac{d^3 k_{l+1}\ldots d^3 k_{n}\, d^3 k_{n+l+1}\ldots 
d^3 k_{2n}}{2^{n-l} (2 \pi)^{3(n-l)}} \nonumber \\ 
& &{}\times \left(\prod_{s=l+1}^{n} \frac{{\mathbf{k}_s}\cdot {\mathbf{k}_{n+s}}}{ \sqrt{\omega_{\mathbf{k}_s}
 \omega_{\mathbf{k}_{n+s}}}} \right)
(-)^{n-l}\langle 0|\prod_{p=l+1}^{n} r_{\mathbf{k}_p}\, r_{\mathbf{k}_{n+p}} |0\rangle  \Bigg ],
 \nonumber
\end{eqnarray}
\begin{eqnarray}
\lefteqn{\langle {\cal H}^n_{\rm C} \rangle =  } \label{vevHcn1ax} \\
&  &{}  \sum_{l=0}^n \left( n \atop l \right)
\Bigg [\int \frac{d^3 k_1\ldots d^3 k_{l}\, d^3 k_{n+1}\ldots d^3 k_{n+l}}{2^l (2 \pi)^{3l}} \nonumber \\
& &{}\times 
\left(\prod_{i=1}^l \sqrt{\omega_{\mathbf{k}_i}
 \omega_{\mathbf{k}_{n+i}}}\right)
 \langle 0|\prod_{j=1}^{l} c^\dagger_{\mathbf{k}_j}\, c_{\mathbf{k}_{n+j}} |0\rangle \Bigg ]
 \nonumber \\
& &{}\times  \Bigg [\int \frac{d^3 k_{l+1}\ldots d^3 k_{n}\, d^3 k_{n+l+1}\ldots d^3 k_{2n}}{2^{n-l} 
(2 \pi)^{3(n-l)}} 
\nonumber \\
& &{}\times
\left(\prod_{s=l+1}^{n} \frac{{\mathbf{k}_s}\cdot {\mathbf{k}_{n+s}}}{ \sqrt{\omega_{\mathbf{k}_s}
 \omega_{\mathbf{k}_{n+s}}}} \right)
\langle 0|\prod_{p=l+1}^{n} c^\dagger_{\mathbf{k}_p}\, c_{\mathbf{k}_{n+p}} |0\rangle  \Bigg ],~
\nonumber
\end{eqnarray}
where, as before, in the case of ill-defined integrals (cases $l=0$ and $l=n$) we assume 
that the contribution of the integral is equal to 1.

Because of the delta functions emerging from matrix elements, integrals in the 
square brackets of Eqs. (\ref{vevHrn1ax}) and (\ref{vevHcn1ax})
reduce to the combinations of the integrals
\begin{eqnarray}
\int \frac{d^3 k}{2 (2 \pi)^3} ~\omega_{\mathbf{k}} &=& \kappa, \label{int1} \\
\int \frac{d^3 k}{2 (2 \pi)^3} ~\frac{k_i k_j}{\omega_{\mathbf{k}}} &=& \frac{\kappa}{3} \, \delta_{ij}, \label{int2}
\end{eqnarray}
where, for further convenience, a short notation ($\kappa$) is introduced for the first integral 
and the second integral is represented in terms of the first integral. 
It is important to note that the above integrals are divergent, so 
is parameter $\kappa$, and 
strictly speaking one has to regularize integrals 
(\ref{int1}) and (\ref{int2}). 
Clearly the physically relevant conclusions should not depend on the choice 
of regularization. However, it is important to discuss various regularizations since, if blindly used, they
can give misleading results. That is especially true for the case of dimensional regularization. Namely,
if dimensional regularization is applied in a usual way the parameter $\kappa$ is put to zero by convention
using the argument that it has no scale dependence. Consequently, that puts all the moments (except the zeroth 
one) of the distribution to zero and the p.d.f. would equal the delta functional. However 
that would imply that the Hamiltonian density and the Hamiltonian commute 
(at least when their commutator acts on the vacuum state) which is not 
the case, as one can check by explicit calculation. To resolve this paradox, one has to remember that
rigorously speaking the integrals (\ref{int1}) and (\ref{int2})
are not zero in dimensional regularization. They are either UV divergent or IR 
divergent depending on the concrete dimension used for regularization. That is because relevant integral
in dimensional regularization can be written in the form $\int_0^{\infty} dk\,k^\alpha$ 
(where $k$ is magnitude of momenta in $D$-dimensional space) which is obviously divergent for any $\alpha$. 
However, for practical purposes (calculation of cross-sections and decay rates) these integrals are conventionally put to zero
because their contributions are always fully absorbed into renormalized quantities and we have complete control over them.
There is not any justification to use such an approach to calculate the p.d.f. of the vacuum energy density. In a sense, the standard
dimensional regularization procedure is not an adequate regularization procedure here. However, since relevant conclusions should
not depend on the choice of the regularization, for our purposes, we can adopt more convenient regularization.
For example, in three-dimensional cutoff regularization where the 
momentum cut-off is $\Lambda$, the parameter $\kappa$ is equal to $\Lambda ^4/(16\,\pi^2)$. 
Anyway, let us note that, 
fortunately, to obtain the 
functional forms of the probability density 
functions it is not necessary to  exactly calculate the above integrals i.e. to know the value of $\kappa$. 
Namely,
 when normalized and expressed in terms of expectation value the p.d.f.s do not depend on 
$\kappa$ explicitly (see Eqs. (\ref{PRx}) and (\ref{PCx}) later in the text).

After integration and counting of identical terms, it can be shown that VEVs 
$\langle {\cal H}^n_{\rm R} \rangle$ and $\langle {\cal H}^n_{\rm C} \rangle$
 are equal to
\begin{eqnarray}
\langle {\cal H}^n_{\rm R} \rangle &=& \frac{1}{2^n} \sum_{l=0}^{n} 
\left( n \atop l \right) \Bigg [ \frac{(2 l)!}{2^l l!}  \kappa^l \Bigg ] \nonumber \\
& & {}\times
 \Bigg [ \left(\frac{\kappa}{3}\right)^{n-l} \sum_{k=0}^{n-l} 3^k\, 2^{n-l-k} 
| S^{(k)}_{n-l} |
\Bigg], \label{Rfin}\\
\langle {\cal H}^n_{\rm C} \rangle &=&  \sum_{l=0}^{n} \left( n \atop l \right)\! \Bigg [  l!\, 
\kappa^l  \Bigg ] \! 
\Bigg [ \left(\frac{\kappa}{3}\right)^{n-l} \sum_{k=0}^{n-l} 3^k\, 
| S^{(k)}_{n-l} |
 \Bigg],\label{Cfin}
\end{eqnarray}
where $| S^{(j)}_i |$ are unsigned Stirling numbers of the first kind. The number $| S^{(j)}_i |$ gives a 
number of permutations of $i$ elements which have exactly $j$ cycles \cite{Abram}. 
In the above equations, terms are still grouped in square brackets to indicate the origin of 
each contribution if compared to Eqs. (\ref{vevHrn1ax}) and (\ref{vevHcn1ax}). 

Now it is possible to calculate Laplace transforms of the probability density functions of the vacuum 
energy for real and complex massless scalar fields. After putting Eqs. (\ref{Rfin}) and (\ref{Cfin}) in
Eq. (\ref{transf}) and rearrangement of terms, 
it follows
\begin{equation}
L\{p_{{\rm R},{\rm C}}(\epsilon)\} = \sum_{n=0}^{\infty}\frac{1}{n!} 
\left( -\frac{s \kappa}{3}\right ) ^n L^{(n)}_{{\rm R},{\rm C}},
\label{lt}
\end{equation}
where
\begin{eqnarray}
L^{(n)}_{{\rm R}} & = & \sum_{l=0}^n 
\left( n \atop l \right)\frac{(2 l)!}{l!}\left(\frac{3}{4}\right)^l
\sum_{k=0}^{n-l}  
| S^{(k)}_{n-l} |
\left(\frac{3}{2}\right)^{k}, \\
L^{(n)}_{{\rm C}} & = & \sum_{l=0}^n  \left( n \atop l \right) l!\, 3^{l} \sum_{k=0}^{n-l} 
| S^{(k)}_{n-l} |
\, 3^{k}.
\end{eqnarray}
Since
\begin{eqnarray}
\left. \frac{d^n}{d\,x^n} (1-x)^{-u}\right |_{x=0}&=&\sum_{k=0}^{n} 
| S^{(k)}_{n} |
\, u^k,\nonumber \\
\left. \frac{d^n}{d\,x^n} (1-3 x)^{-1/2}\right |_{x=0}&=& \frac{(2 n)!}{n!}\left(\frac{3}{4}\right)^n,\nonumber \\
\left. \frac{d^n}{d\,x^n} (1-3 x)^{-1}\right |_{x=0}&=& n!\,3^n,\nonumber
\end{eqnarray}
it can be seen that functions $L^{(n)}_{{\rm R},\,{\rm C}}$ equal
\begin{eqnarray}
L^{(n)}_{\rm R} & = & \left. \frac{d^n}{d\,x^n} \frac{1}{(1-x)^{3/2}(1-3 x)^{1/2}}\right |_{x=0}, \label{GR}\\
L^{(n)}_{\rm C} & = & \left. \frac{d^n}{d\,x^n} \frac{1}{(1-x)^3(1-3 x)}\right |_{x=0}.\label{GC}
\end{eqnarray}
Finally, by taking into account identities (\ref{GR}), (\ref{GC}), and (\ref{lt}), the relevant Laplace 
transforms are
\begin{eqnarray}
L\{p_{\rm R}(\epsilon)\} & = & (1+\kappa s/3)^{-3/2}(1+\kappa s)^{-1/2}, \label{GRf}\\
L\{p_{\rm C}(\epsilon)\} & = & (1+\kappa s/3)^{-3}(1+\kappa s)^{-1}.\label{GCf}
\end{eqnarray}

By using the differentiation theorem of Laplace transform, it further follows
\begin{eqnarray}
L\{p_{\rm R}(\epsilon)\} & = & \frac{6}{\kappa} \left(-\frac{d}{d\,s}
(1+\kappa s/3)^{-1/2}\right)(1+\kappa s)^{-1/2} \nonumber \\ 
& = & \frac{6}{\kappa} L\{\epsilon f(\epsilon,\kappa/3)\}L\{f(\epsilon,\kappa)\},
\end{eqnarray}
where
$L\{f(\epsilon,a)\}=(1+a s)^{-1/2}$. Since
\[L\{\frac{e^{-t/a}}{\sqrt{\pi a t}}\}=\frac{1}{\sqrt{1+a s}},\]
it can be written
\begin{eqnarray}
L\{p_{\rm R}(\epsilon)\} 
& = & \frac{6}{\kappa} L\{\frac{\sqrt{3\, \epsilon}e^{-3\epsilon/\kappa}}{\sqrt{\pi \kappa}}\}
L\{\frac{e^{-\epsilon/\kappa}}{\sqrt{\pi \kappa \epsilon}}\}  \\ 
& = & L\{\frac{6 \sqrt{3}\, \epsilon}{\pi \kappa^2}e^{-3\epsilon/\kappa}\int_0^1 dx \,
\sqrt{\frac{1-x}{x}}e^{2\epsilon x/\kappa}\}, \nonumber
\end{eqnarray}
where theorems of linearity and convolution are used. After calculating the remaining integral we obtain
\begin{eqnarray}
\label{pr}
p_{\rm R}(\epsilon)= \frac{3 \sqrt{3}\,\epsilon}{\kappa^2} e^{-2 \epsilon/\kappa}\left[ I_0\left(\frac{\epsilon}{\kappa}\right)-I_1\left(\frac{\epsilon}{\kappa}\right)\right],
\end{eqnarray}
where $I_n(x)$ is a modified Bessel function of the first kind.

Similarly, for the complex scalar field
\begin{eqnarray}
L\{p_{\rm C}(\epsilon)\} & = & L\{\frac{27\,\epsilon^2}{2\, \kappa^3} e^{-3\,\epsilon/\kappa}\}
L\{\frac{1}{\kappa} e^{-\epsilon/\kappa}\},\label{bb}
\end{eqnarray}
where the following identity is used,
\[L\{\frac{t^{n-1}}{a^n(n-1)!}e^{-t/a}\}=\frac{1}{(1+a s)^n}.\]
Applying the convolution theorem to Eq.(\ref{bb}) gives
\begin{eqnarray}
L\{p_{\rm C}(\epsilon)\} & = & L\{\frac{27\,\epsilon^3}{2\, \kappa^4} 
e^{-\epsilon/\kappa} \int_0^1 dx\,x^2 e^{2\epsilon x/\kappa}\},
\end{eqnarray}
and finally after integration we get
\begin{equation}
\label{pc}
p_{\rm C}(\epsilon)= \frac{27}{8 \kappa}\left\{
e^{-\epsilon/\kappa}-e^{-3\,\epsilon /\kappa}\left[
 2 \left(\frac{\epsilon}{\kappa}\right)^2+2
\frac{\epsilon}{\kappa}+1\right]\right\}.
\end{equation}
It is convenient to express the probabilities 
$p_{{\rm R},{\rm C}}(\epsilon)\, d\epsilon$ in terms of vacuum 
expectation values $\bar{\epsilon}_{{\rm R},{\rm C}}=
\langle {\cal H}_{{\rm R},{\rm C}} \rangle $ as
\begin{equation}
p_{{\rm R},{\rm C}}(\epsilon)\,d\epsilon= P_{{\rm R},{\rm C}}
\left(\frac{\epsilon}{\bar{\epsilon}_{{\rm R},{\rm C}}}\right)
\,d\left(\frac{\epsilon}{\bar{\epsilon}_{{\rm R},{\rm C}}}\right),
\end{equation}
where
$\bar{\epsilon}_{\rm R}=\kappa$, $\bar{\epsilon}_{\rm C}=2\,\kappa$,
and
\begin{eqnarray}
P_{\rm R}(x)&=& 3 \sqrt{3}\, x\, e^{-2 x}\left[ I_0\left(x\right)-I_1\left(x\right)\right], \label{PRx}\\
P_{\rm C}(x)&=& 27/4  \left[e^{-2 x}-e^{-6 x}(1+4 x+8 x^2)\right]. \label{PCx}
\end{eqnarray}

\begin{figure}
\begin{center}
\includegraphics[width=8cm]{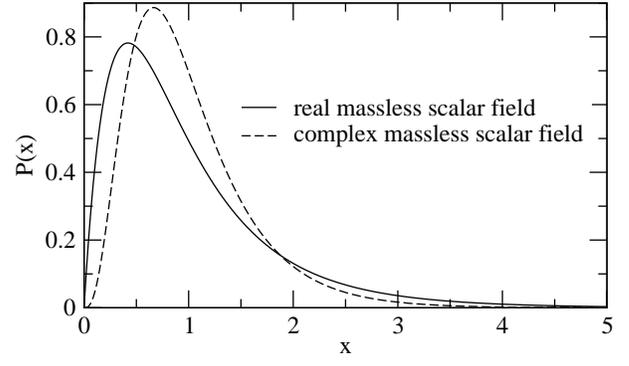}
\end{center}
\caption{\label{slika}
The probability density functions of the vacuum energy density for real (solid) and 
complex (dashed) massless scalar fields. Variable $x$ is equal to the ratio 
$\epsilon/\bar{\epsilon}$ where $\epsilon$ is the energy density and 
$\bar{\epsilon}$ is the expectation value which is different for the real 
and the complex cases. Mean values for both curves are $\bar{x}=1$ and the 
standard deviation for the real case is $\sigma=\sqrt{2/3}=0.816$ and 
for the complex case is $\sigma=\sqrt{1/3}=0.577$. For the real field 
maximal probability is at $x=0.419$ and for the complex field at $x=0.664$.}
\end{figure}

As can be seen from  Fig. \ref{slika}, probability density functions 
$P_{{\rm R},\,{\rm C}}(x)$ are broad.

\begin{figure}
\begin{center}
\includegraphics[width=8cm]{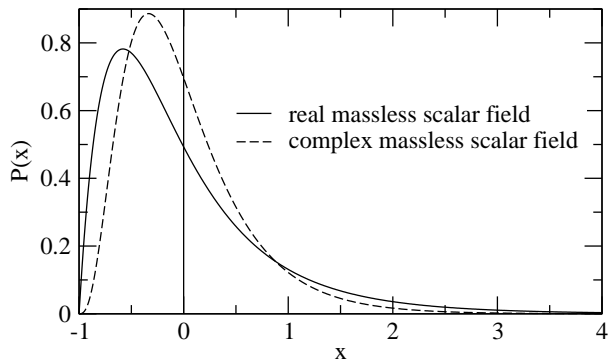}
\end{center}
\caption{\label{slika1}
The probability density functions of the normally ordered vacuum energy density for real (solid) and 
complex (dashed) massless scalar fields. Variable $x$ is equal to the ratio 
$\epsilon/\kappa$ for the real and $\epsilon/2 \kappa$ for the complex field.
Parameter $\epsilon$ is the energy density and 
$\kappa=\Lambda ^4/(16\,\pi^2)$  in three-dimensional cutoff regularization, where the 
momentum cutoff is $\Lambda$.}
\end{figure}

The distributions given in
(\ref{pr}) and (\ref{pc}) depend on formally divergent quantity
$\kappa$
and a regularization followed by renormalization must be employed. 
The only required renormalization  on free field theories
within the standard approach in QFT is given by normal ordering. As
normal ordering corresponds to subtraction of $\langle {\cal H} \rangle$
from ${\cal H}$, the effect of this renormalization on probability
distributions is their shift so that their expectation values vanish as shown in  Fig. \ref{slika1}.
This can be checked by direct calculation if $:{\cal H}:={\cal H}-\langle {\cal H} \rangle$
is used instead of ${\cal H}$, which requires calculation of $\langle :{\cal H}:^n \rangle$ in intermediate
steps.
However, the form of 
probability distributions 
 for $:{\cal H}:$
still depends on 
formally divergent
$\kappa$.
In a sense, this
indicates that the normal
ordering is insufficient to remove all divergences from the
distributions.
However, it is not realistic that $\kappa$ is really divergent. Any plausible energy cut-off makes it big but finite.
All moments can be naively renormalized if moments are defined as $\langle :{\cal H}^n: \rangle$
instead of $\langle {\cal H}^n \rangle$ or $\langle :{\cal H}:^n \rangle$. 
Here, it is important to note that such procedure corresponds to infinitely independent renormalizations
of moments since each moment has a different divergency (beside the fact they all
depend on the same formally divergent integral $\kappa$).
With such an approach, all the moments (except the zeroth 
one) would be equal zero and the p.d.f. would equal the delta functional, which leads us into contradiction
with the fact that the Hamiltonian density and Hamiltonian do not commute as mentioned above.

In conclusion, the calculations presented in this paper show that the vacuum energy density of a
 massless scalar field is a random variable with a broad p.d.f. Both the expectation of 
the vacuum energy density and the most probable value of the vacuum energy density 
are not representative for the distribution as a whole. There exists a nonnegligible 
probability for a value of vacuum energy that differs a lot from $\overline{\epsilon}$. 
These results, obtained analytically and in a closed form, imply that the vacuum energy 
density is a more complex object than just the 
VEV of ${\cal H}$.

The central results of this work have implications beyond the scope of this paper. 
A thorough understanding of the vacuum energy density is essential for the explanation 
of the epochs of accelerated expansion such as the inflationary phase or the late-time 
accelerated expansion. An especially interesting prospect is that a multiverse scenario 
could be realized in the framework of QFT. To this end, the present analysis needs to 
be generalized to curved space-times to study the zero point energy density in the presence of 
gravity. Natural extensions of the present analysis comprise the study of effects of 
mass, other types of quantum fields, and field interactions. The analytical results 
presented here represent a sound basis for the exploration of these directions of research.  

This work was supported by the Ministry of Education,
Science and Sports of the Republic of Croatia under Contract No.
098-0982930-2864.


\end{document}